\documentclass[5p]{elsarticle}
\usepackage{graphicx}
\usepackage[usenames, dvipsnames]{color}
\usepackage{dcolumn}
\usepackage{bm}
\usepackage{amsmath}
\usepackage{amssymb}
\usepackage{txfonts}
\usepackage{float}

\newcommand{\ket}{\rangle}

\def\be{\begin{equation}}
\def\ee{\end{equation}}
\def\bea{\begin{eqnarray}}
\def\eea{\end{eqnarray}}
\def\bse{\begin{subequations}}
\def\ese{\end{subequations}}
\def\br{{\bf r}}

\def\avr#1{\left\langle{#1}\right\rangle}
\def\ket#1{\left|{#1}\right\rangle}

\def\brakket#1#2#3{\left\langle{#1}\middle|{#2}\middle|{#3}\right\rangle}
\usepackage[T1]{fontenc}

\begin{document}

\title{Isovector density and isospin impurity in $^{40} \mathrm{Ca}$}


 \author{H.~Sagawa\fnref{1,2}}
\address[1]{RIKEN Nishina Center for Accelerator-Based Science, Wako, Saitama 351-0198, Japan}
\address[2]{Center for Mathematical Sciences, the University of Aizu, Aizu-Wakamatsu, Fukushima 965-8580, Japan}

\author{S.~Yoshida\fnref{3}}
\address[3]{Science Research Center, Hosei University,
2-17-1 Fujimi, Chiyoda, Tokyo 102-8160, Japan}

\author{T.~Naito\fnref{1,4}} 
  \address[4]{Department of Physics, Graduate School of Science, The University of Tokyo, Tokyo 113-0033, Japan}

\author{T.~Uesaka\fnref{1,5}}
\address[5]{RIKEN Cluster for Pioneering Research, Wako, Saitama 351-0198, Japan}

 \author{J.~Zenihiro\fnref{1,6}}
\address[6]{Department of Physics, Kyoto University, Kitashirakawa-Oiwake, Sakyo, Kyoto 606-8502, Japan}

 \author{J.~Tanaka\fnref{1}}
 \author{T. Suzuki\fnref{7}}
\address[7]{Department of Physics, College of Humanities and Sciences, Nihon University, Sakurajosui 3, Setagaya-ku, Tokyo 156-8550, Japan} 
\vspace{-1cm}

  
\begin{abstract}
We study 
isoscalar (IS) and  isovector (IV) densities in  ${}^{40} \mathrm{Ca}$ in comparison with theoretical densities calculated by Skyrme Hatree-Fock (HF) 
  models.
The charge symmetry breaking  and the charge independence breaking forces are introduced to
study the effect on the IV density.
The effect of isospin mixing in the ground-state density is examined by using the particle-vibration coupling model 
taking into account the  collective IV giant monopole excitation.
We show a  clear correlations  in the  IV density and isospin impurity of ${}^{40} \mathrm{Ca}$ within the HF and the particle-vibration coupling model.
We extract for the first time the experimental information of isospin impurity from the magnitude of IV density.
 \end{abstract}
\maketitle
\section{Introduction}

Experimental charge densities have been studied by electron scattering experiments from 1950s. 
Proton densities can be extracted from the charge densities removing the proton finite-size effect.
It has been shown in Refs.~\cite{Star1994,Zenihiro,Zenihiro2} that the  proton elastic scattering  is quite useful to extract the matter distributions of nuclear ground states.
Combining the proton and the matter density distributions, the neutron density can be extracted from the  experimental data.   
The neutron density distributions and neutron-skin thicknesses, $ \Delta r_{np} $, 
in ${}^{40} \mathrm{Ca}$ and ${}^{48} \mathrm{Ca}$ are recently determined from the angular distributions
of the cross sections and the analyzing powers of  polarized proton
elastic scattering at $E_p = 295 \, \mathrm{MeV}$ \cite{Zenihiro}.  
Experimental proton and neutron rms radii are listed for ${}^{40} \mathrm{Ca}$ and ${}^{48} \mathrm{Ca}$ in Table \ref{tab1} together with those of ${}^{208} \mathrm{Pb}$.  
The determination of the proton and neutron density distributions and
$\Delta r_{np}$ of ${}^{48} \mathrm{Ca}$ gives a unique opportunity to examine more comprehensively nuclear and neutron matter EoS at various densities \cite{Zenihiro2021}.
 

The Skyrme 
HF model is one of the most successful mean-field models to describe  the ground-state properties including   single-particle energy spectra of  closed-shell and  open-shell  nuclei. 
These models are applied also to describe excited states  such as low-lying collective states and giant resonances. 
 The Skyrme interactions 
  were originally determined by fitting 
 charge radii and masses of closed shell nuclei,  while some parameter sets were optimized including the nuclear matter incompressibility, the Landau-Migdal parameters and/or the  neutron matter EoS in the input data. 
While these effective interactions can reproduce  the 
saturation properties of symmetric nuclear matter well,  they  have generally different 
isoscalar and isovector nuclear matter properties.
We use hereafter modern version of energy density functions (EDFs) \cite{Yoshida2020};  SAMi-J family as  Skyrme EDFs 
 for the study of IS and IV densities. 

The charge symmetry breaking (CSB) and charge independence breaking (CIB) forces  have been discussed in the context of isospin impurity effect on the super-allowed Fermi decays.
The quantitative information of isospin symmetry breaking (ISB) forces is recently examined 
to calculate the binding energies of isodoublet and isotriplet nuclei \cite{Jacheck} and also the excitation energies of isobaric analogue states (IAS)~\cite{Xavi}.  The ISB interactions are not included in  the standard Skyrme EDFs.  
In this paper, we introduce the Skyrme-type ISB interactions and study how the IV density in a $ N=Z $ nucleus ${}^{40} \mathrm{Ca}$  is affected by these interactions.
Furthermore, we will extract the correlation between IV density and the isospin impurity through the ISB forces.
The IV density is also examined by using  the particle-vibration coupling (PVC) model taking into account the IV giant monopole resonance, which will
be useful to relate between the isospin impurity and the IV density.

\begin{table}[tb]
  \caption{
    Table of  empirical rms radii and  neutron-skin thicknesses.
    $r_{\mathrm{ch}}$ and $r_p$, 
    and the extracted $r_n$ and $\Delta r_{np}$
    in ${}^{40} \mathrm{Ca}$,  ${}^{48} \mathrm{Ca}$ and ${}^{208} \mathrm{Pb}$ are listed.
     Experimental data are taken from Refs.~\cite{Zenihiro,Zenihiro2,electron-sca}. All data are written in unit of fm.}
    \label{tab1}
    \begin{tabular}{llD{.}{.}{4}D{.}{.}{4}D{.}{.}{4}D{.}{.}{4}}
      & Exp. & \multicolumn{1}{c}{$r_{\mathrm{ch}}$} & \multicolumn{1}{c}{$r_p$} & \multicolumn{1}{c}{$r_n$} & \multicolumn{1}{c}{$\Delta r_{np}$}
      \\ \hline
      ${}^{40} \mathrm{Ca}$
        & 
        Ref.~\cite{Zenihiro,electron-sca}  & 3.480 & 3.385 & 3.375 & -0.010 \\
      ${}^{48} \mathrm{Ca}$
        & 
        Ref.~\cite{Zenihiro,electron-sca}  & 3.460 & 3.387 & 3.555 & 0.168  \\
      ${}^{208} \mathrm{Pb}$ & 
      Ref.~\cite{Zenihiro2,electron-sca} & 5.503& 5.442& 5.653&0.211  \\
    \end{tabular}
   \vspace*{-0.5cm}  
\end{table}

\begin{figure}[tb]
  \vspace*{2.0cm}
  \includegraphics[width=0.9\linewidth,
  bb=0 250 482 562]{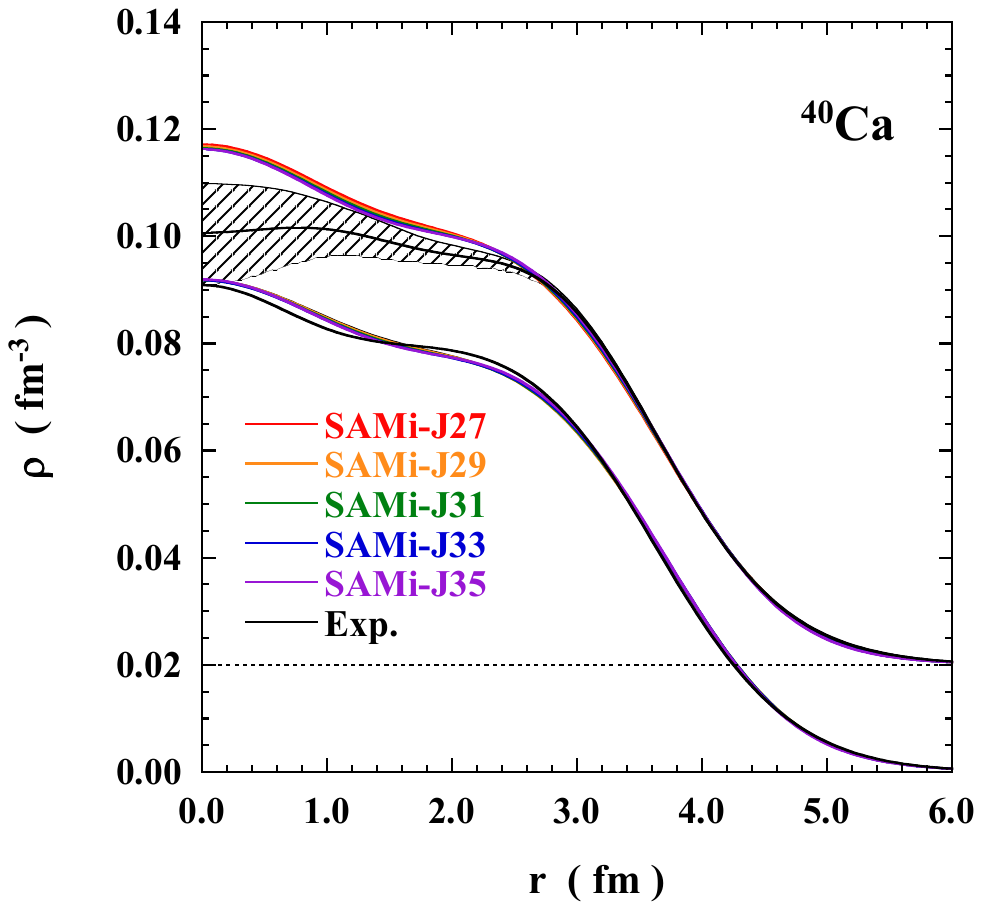}
 \vspace*{-0.3cm}
\caption{(color online) 
Experimental proton and neutron densities of ${}^{40} \mathrm{Ca}$ together with calculated ones using  SAMi-J  interactions. 
For a guide to eyes, the neutron density is shifted by 0.02 fm$^{-3}$.  
 The black solid lines show experimental data  taken from Ref.~\cite{electron-sca} for protons and from Ref.~\cite{Zenihiro} for neutrons. 
 The shaded area of experimental neutron density shows experimental uncertainties of statistical and systematic errors.
 \label{fig2}}
 \vspace*{-0.3cm}
\end{figure}

\begin{figure}[tb]
 \vspace*{2.0cm}
  \includegraphics[width=0.9\linewidth,bb=0 300 482 612]{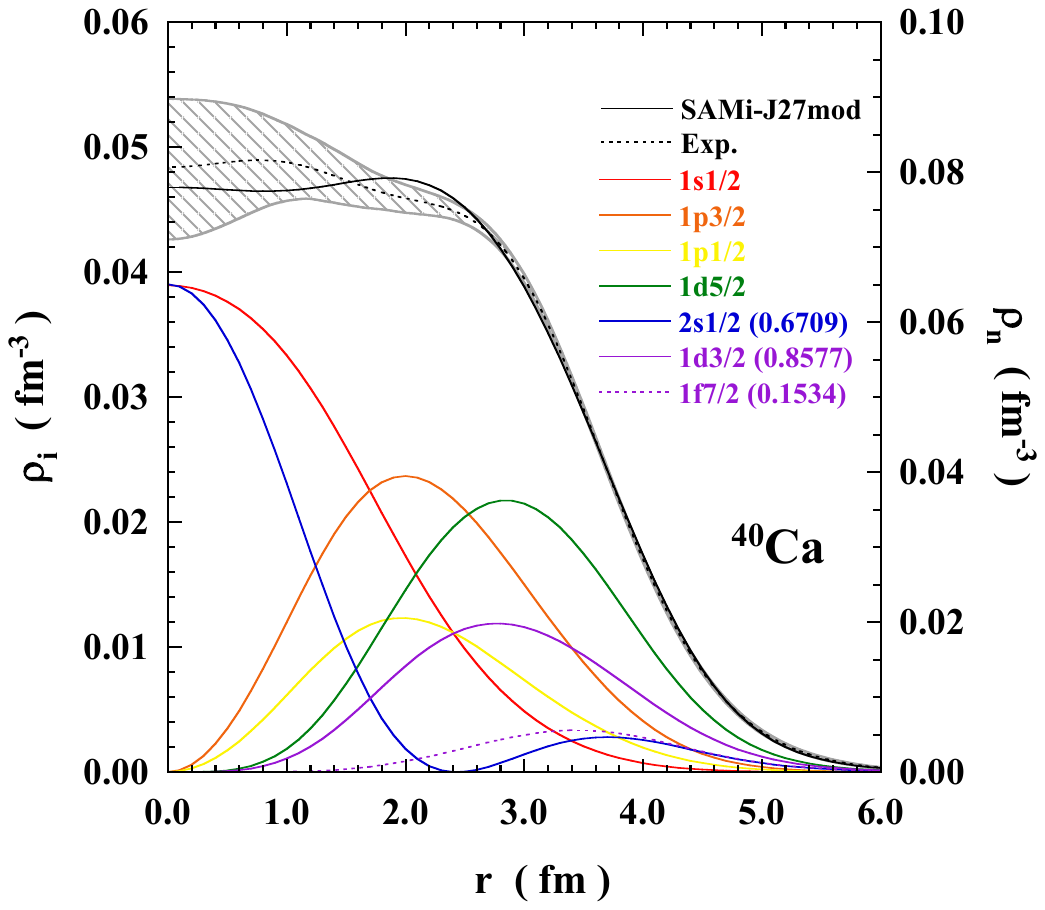}
 \vspace*{-0.3cm}
\caption{(color online) Calculated neutron density of ${}^{40} \mathrm{Ca}$ 
with modified occupation probabilities of  orbits around the Fermi energy.   The contribution of each orbit  $\rho_i$ with a modified  occupation probability, $v_i^2$, (written in the bracket) for the neutron density is shown
with different color.   The modified neutron density is denoted as SAMi-J27mod.  
The  orbits without occupation probabilities are fully occupied.  
See the text for details. 
\label{fig2a}}
 \vspace*{-0.3cm}
\end{figure}

%
\section{Density distributions of  ${}^{40} \mathrm{Ca}$}

Experimental neutron and proton densities of ${}^{40} \mathrm{Ca}$ are shown in Fig. 
  \ref{fig2} together with HF results using 
  SAMi-J families. 
For a guide to eyes, the neutron density is plotted
shifting upwards by a unit of 0.02 fm$^{-3}$.   Experimental proton density is deduced from the charge density observed by electron scattering  subtracting the contribution of finite proton size \cite{electron-sca,Zenihiro}.
The observed neutron density has a large uncertainty as shown by a shaded area in these figures, while the proton density is better determined experimentally with a small uncertainty.  
  It is clear that the determination of density distribution is more uncertain in the interior region compared with the surface region. 
  
  In Fig.~\ref{fig2},  
   we can see 
    a small variation  in the calculated densities 
    depending on   different SAMi-J interactions.  
  The SAMi-J family has  a variation in the symmetry energy coefficient $J$ from $27$ to $35 \, \mathrm{MeV}$ as well as other  symmetry energy coefficients $L$ and $K_{\mathrm{sym}}$.  Calculated nuclear matter properties and rms radii by  SAMi-J families are listed in the supplemental materials.  
In general, the SAMI-J  model underpredicts both proton and neutron densities at around a half  of the saturation density near $r\simeq 3.0 \, \mathrm{fm} $,  
   while the model overpredicts in the interior region $ 0 \, \mathrm{fm} \le r \le 3 (1.5) \, \mathrm{fm}$ for neutrons (protons). 
   In 
  Fig. 
     \ref{fig2},   there are small, but systematic differences in the neutron-skin $\Delta r_{np}$ as listed in Table 1 in the supplemental materials:
   a smaller $J$ value as well as $L$ value produces a larger negative neutron-skin. The calculated results of relativistic mean field model with DDME-J Lagrangians are also shown in the supplemental materials.  General features of calculated results are quite similar to those of the Skyrme model, but the deviations of calculated results from empirical data are somewhat larger  than those of the Skyrme models.

   Figure \ref{fig2a} shows the calculated neutron density of  ${}^{40} \mathrm{Ca}$ with modified occupation probabilities of  orbits around the Fermi energy.
   In the HF calculations,  all $sd$-shell orbits are fully occupied, but the $f$-shell orbits are empty.
In Fig.~\ref{fig2a},   the occupation probabilities of $2s_{1/2}$ orbit and $1d_{3/2}$ orbit are reduced to be 0.67
and 0.86, respectively, while that for $1f_{7/2}$ orbit is increased to be 0.15.  The modification is rather arbitrary, but intended to decrease the central part of density and  to increase the surface region.
In Fig.~\ref{fig2a}, as on purpose,  the agreement  between the experimental and calculated densities becomes better than that  in Fig.~\ref{fig2}, where one can see a clear difference between the experiment and the calculations in the interior part of the density.  This phenomenological approach tells us that the correlations beyond the mean field might be manifested in the neutron density, 
 especially in the interior part of nucleus.  
 
 Large scale shell model calculations have been performed 
  including  the $pf$-shell model configurations in $^{40}$Ca (see the supplemental materials for details). 
  We calculate also the particle occupation numbers in $sd$-$pf$ shell orbits including full two major shell configurations. We found that 
  the summed occupation numbers $\sum_{j=p,f}(2j+1)v_i^2$ in $pf$-shell to be   0.7 including up to  4-particle-4-hole configurations from the $sd$-shell closed core.
  This value is consistent with the experimental analysis of proton transfer reactions on $^{40}$Ca \cite{p-trans}.  In Fig. \ref{fig2a}, the occupation number of $2s_{1/2}$ orbit is crucial to decrease the central part of neutron density.  The empirical value 
  from Fig.  \ref{fig2a} is $(2j+1)v_i^2=1.34$, which is smaller than the present shell model value 1.85.  A smaller particle occupation number for $2s_{1/2}$, 1.7,  was 
  suggested also in the experimental analysis of Ref. \cite{p-trans}.  On the contrary, the occupation number of $1f_{7/2}$ orbit is large as 1.2 in Fig. \ref{fig2a}, while the shell models give about 0.7. 
   The small occupation probability of $2s_{1/2}$ is an interesting open question to be addressed in future study.

\begin{figure}
\begin{center}
\includegraphics[clip,width=1.0\linewidth,bb=20 300 505 742]{
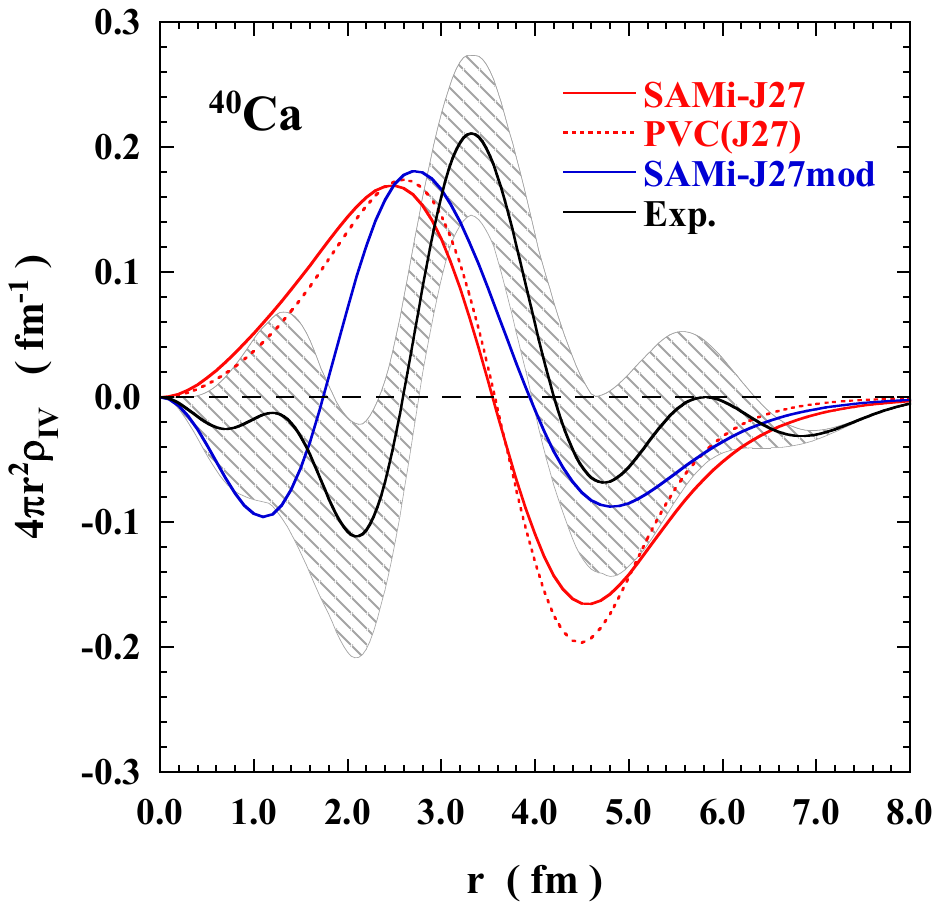}
\vspace{-.8cm}

\caption{(color online) IV densities multiplied by a factor $4\pi r^2\rho_{\rm IV}$ of ${}^{40} \mathrm{Ca}$ calculated by HF, modified HF and PVC models as well as the experimental IV density.
The red (blue) solid and dashed curves show the HF (modified HF) and PVC densities with SAMi-J27 (SAMi-J27mod) model, respectively.
The experimental data are shown with a black solid curve with the experimental uncertainty shown by the shaded area.
The experimental data is taken from Ref.~\cite{Zenihiro}, while the calculated densities are obtained by using the HF and modified ground state densities denoted SAMi-J27 and SAMi-J27mod  in Figs.~\ref{fig2} and \ref{fig2a}, respectively.
See the text for details. \label{fig:PVC-den}}
\vspace{-.8cm}
\end{center}
\end{figure}

In Fig.~\ref{fig:PVC-den}, the isovector density defined by a difference between the neutron
		and proton densities as $\rho_{\rm IV}= \rho_n - \rho_p$ is plotted multiplied by  a phase 
		space factor $4\pi r^2$. In spite of the overall success of the mean field theories
		in reproduction of experimental neutron and proton densities, the theoretical 
		predictions of the isovector density are qualitatively different from the experimental 
		one: the experimental result of the isovector density has a peak at around 
		$r\sim3.2$ fm,  while the HF model with SAMi-J27 interaction predicts a peak at $r\sim$2.5 fm with positive values (neutron excess)
		in the interior and negative values (proton excess) at the surface region.
		The theoretical predictions can be intuitively understood as swelling of the proton
		distribution due to the repulsive Coulomb force.
		In the following section, we will theoretically investigate the behavior of the 
		isovector density from a viewpoint of the isospin impurity in a nucleus. 
		
\section{A particle-vibration coupling model and IV density}

 The isospin impurity can be evaluated by using HF,  and PVC  models to IV giant monopole resonance (GMR).
 We will adopt hereafter the PVC model to illustrate analytically the connection between the isospin impurity and the IV density.
 In the PVC model for the  evaluation of isospin impurity,   the ground state is calculated firstly without the Coulomb and ISB interactions.  Then the ISB forces are introduced by  the first-order perturbation as
\be  \label{eq3}
\ket{\widetilde{\rm GS}} \simeq \ket{{\rm GS}} + \varepsilon^{\tau=1} \ket{{\rm GMR,  \tau=1}},
\ee
taking into account the coupling between the ground state and the collective  IV GMR.  
Here,  
 $\tau=1$ denotes the IV GMR.  
The coefficient $\varepsilon^{\tau=1}$ of the perturbed state is calculated as
\be   \label{eq4}
\varepsilon^{\tau=1}=\frac{\brakket{\rm  GMR, \tau=1 }{V^{\rm ISB}}{\rm GS}}{\Delta E^{\tau=1}},
\ee
where the energy difference in the denominator is given by $\Delta E^{\tau=1}=E_{\rm GS}-E_{\rm GMR}^{\tau=1}$ 
for the  IVGMR. 
 In the numerator, 
  $V^{\rm ISB}$  is
the isospin symmetry breaking interactions. 
 
The perturbed IV density for the  ground state  is then expressed as
\begin{align}
  & \brakket{\widetilde{\rm GS}}{\hat{\rho}_{\rm IV}(\br)}{\widetilde{\rm GS}}
    \nonumber \\
  & = 
    \brakket{\rm GS}{\hat{\rho}_{\rm IV}(\br)}{\rm GS}
    +
    2\varepsilon^{\tau=1}
    \brakket{{\rm GMR},\tau=1}{\hat{\rho}_{\rm IV}(\br)}{\rm GS}
    \nonumber \\
  & = 
    \rho_{\rm IV}(r)+2\varepsilon ^{\tau=1} \rho^{\rm IVTR}_{\tau=1}(\br)
  \equiv
    \rho_{\rm IV}(r)+\Delta \rho^{\tau=1}_{\rm IVTR},
  \label{eq5}
\end{align}
where the IV density operator reads $\hat{\rho}_{\rm IV}(\br)=\sum_i \delta (\br -\br_i)\tau_{z}(i)$ and 
$\rho^{\rm IVTR}_{\tau}(\br)$ is the transition density defined by
\be
\rho^{\rm IVTR}_{\tau=1}(\br)\equiv \brakket{{\rm GMR}, \tau=1}{\hat{\rho}_{\rm IV}(\br)}{{\rm GS}}.
\ee
Notice that the first term of Eq.~\eqref{eq5} disappears for $N=Z$ nuclei without the ISB interactions.  
Under the assumption that one collective IV GMR exhausts fully the sum rule strength, the GMR transition density is expressed by
the Werntz-\"Uberall type
  \cite{WU66}, 
\be   \label{eq7}
\rho_{\tau=1}^{\rm IVTR}(\br)=\alpha^{\tau=1}\rho^{\rm WU}_{\rm tr}\equiv\alpha^{\tau=1} \frac{1}{r^2} \frac{d}{dr}\left(r^3\rho_{\rm IS}(r)\right) Y_{00}(\hat{r}).
 \ee
where  the amplitude $\alpha^{\tau=1}$ determines   the collectiveness of the GMR excitation,  $\rho_{\rm IS}(r)$ is the IS ground state density for the operator $\hat{\rho}_{\rm IS}(\br)=\sum_i\delta(\br-\br_i)$,
 and  $Y_{00}(\hat{r})$ is the spherical harmonics with the multipole $\lambda=0$.  A similar  radial form of transition density is  obtained by a hydrodynamical model of compressive irrotational fluid \cite{BM2}.
 The collective parameter $\alpha^{\tau=1}$  
  is normalized to satisfy the energy-weighted sum rule (EWSR) $m_1$ for the IVGMR,
\bea  \label{eq6}
m_1^{\tau=1}&\equiv& \sum_n \hbar \omega^{\tau=1}_n \left| \brakket{n,\tau=1}{\hat{O}_{\tau=1}^{\lambda=0}}{{\rm GS}} \right|^2  \nonumber \\
   &=&\frac{2\hbar^2}{m}\frac{A}{4\pi}\avr{r^2}_m,  
   \eea
where 
the monopole transition operator is given by 
\bea
 \hat{O}_{\tau=1}^{\lambda=0}=\sum_i r_i^2 Y_{00}(\hat{r})  \tau_{z}(i).   
 \eea
In Eq. \eqref{eq6}, 
  $\avr{r^2}_m$ is the matter mean square radius and $\hbar \omega^{\tau=1}_n=E^{\tau=1}_n-E_{\rm GS}$.  
  The enhancement of the sum rule  by exchange interactions for the IV excitation is neglected in Eq.~\eqref{eq6}.
  The IV EWSR is also expressed by using the transition density as
\be 
    m_1^{\tau=1}
 =    \hbar \omega_{\rm GMR}^{\tau=1}
      \left| \alpha^{\tau=1} \right|^2
      \left|
      \int \rho^{\rm WU}_{\rm tr}r^4 \, dr
      \right|^2
 =      \hbar \omega_{\rm GMR}^{\tau=1}
      \left| \alpha^{\tau=1} \right|^2
      \frac{A^2}{4\pi^2} \avr{r^2}_m^2,    
 \ee 
where we assume that  the single collective IV GMR state 
 exhausts the EWSR having  the excitation energy $\hbar \omega_{\rm GMR}^{\tau=1}$. 
 The collective amplitude is then obtained as 
\bea \label{eq:alpha}
       \left|\alpha^{\tau=1}\right|^2=\frac{2\hbar^2}{m}\frac{\pi}{\hbar \omega_{\rm GMR}^{\tau=1} A\avr{r^2}_m}  
=\frac{301.5}{\hbar \omega_{\rm GMR}^{\tau=1}}A^{-5/3},
\eea
where the mean square radius is taken as $\avr{r^2}_m=3r_0^2A^{2/3}/5$ fm$^2$ with $r_0=1.2$ fm.  The  effect of PVC of IVGMR on the IV density reads
\be  \label{eq9}
\Delta \rho_{\rm IVTR}^{\tau=1}(\br)=2\varepsilon ^{\tau=1} \rho^{\rm IVTR}_{\tau=1}(\br)=2\varepsilon ^{\tau=1}\alpha^{\tau=1} \frac{1}{r^2} \frac{d}{dr}\left(r^3\rho_{\rm IS}(r)\right) \frac{1}{\sqrt{4\pi}}.  
\ee

\section{Isospin impurity and isospin symmetry breaking forces}
 
 
 In the study of isospin impurity of nuclear ground state,  the coupling of IVGMR was introduced  to evaluate 
 the  isospin impurity in the unperturbed ground state with the good isospin \cite{Auerbach}. 
  The IVGMR is mixed by the isospin symmetry breaking interaction $V^{\rm ISB}$ as written in Eq.~\eqref{eq3}.   
 In the following, we consider only the Coulomb force to break the isospin symmetry.  The effect of CSB and CIB interactions will be also discussed later. 

The mixing amplitude $\varepsilon^{\tau=1}$ might be  evaluated with the Coulomb potential,
\be
V_C(r)=\frac{Ze^2}{R}\left(\frac{3}{2}-\frac{r^2}{2R^2}\right),
\ee
which is obtained by assuming a constant charge distribution in the sphere of radius $R$.
The mixing matrix element is then expressed as
\be
\brakket{\rm GMR, \tau=1}{V^{\rm ISB}}{\rm GS}=\int d\br \rho^{\rm IVTR}_{\tau=1}(\br)V_C(r)
\ee

In the proton-neutron two-fluid model \cite{BM2}, the excitation energy of IVGMR is estimated to be
\be   \label{BM-IVGMR} 
\hbar \omega^{\tau=1}_{\rm GMR}=170/A^{1/3} \, \mathrm{MeV},
\ee
which is 49.7MeV for ${}^{40} \mathrm{Ca}$.  Taking the Werntz-\"Uberall-type transition density for IVGMR having the full sum rule strength and the IVGMS energy \eqref{BM-IVGMR}, the isospin impurity is evaluated as
\be \label{IS-impurity}
\varepsilon^2=2.5\times 10^{-6}Z^4/A^{4/3},
\ee
which gives $\varepsilon^2$=0.29\% for ${}^{40} \mathrm{Ca}$.  This value is about two times larger than the hydrodynamical  model evaluation in Ref.~\cite{BM1},
$\varepsilon$ (two-fluid)$^2=3.50\times 10^{-7}Z^2A^{2/3}$. 
The self-consistent RPA calculations for IVGMR were performed in Ref.~\cite{HSZ}
with Skyrme interactions.  With a Skyrme interaction SIII, the IV monopole strength is rather widely spread in energy and the average excitation energy in ${}^{40} \mathrm{Ca}$ becomes  a much lower energy,   $\overline{\hbar \omega}^{\tau=1}=33.5 \, \mathrm{MeV} $,   than Eq.~\eqref{BM-IVGMR}.     Consequently,  the microscopic HF+RPA model gives even larger isospin mixing  than that given in Eq.~\eqref{IS-impurity}.  In Ref.~\cite{HS}, the isospin impurity is estimated as 0.57 \% in ${}^{40} \mathrm{Ca}$  by using HF+TDA model with SIII Skyrme interaction.   The sum rule value of Fermi transitions is
linked to the isospin impurity in the HF+TDA model, while the present HF model calculates directly the overlap between the proton and neutron HF wave functions to evaluate the isospin impurity.  The two models are equivalent to evaluate the impurity, while the HF+RPA model provides an additional effect of the
ground state correlations on the isospin impurity.

 For the collective amplitude of IVGMR in ${}^{40} \mathrm{Ca}$, Eq.~\eqref{eq:alpha} gives $|\alpha^{\tau=1}|^2=0.0192$ for the IVGMR  energy with SIII interaction.  With these $\alpha^{\tau=1}$ and $\varepsilon$, the renormalized  transition density gives the IV density in the ground state and compared with the HF isovector density in Fig.~\ref{fig:PVC-den}.  It is remarkable that 
the PVC  density obtained by using SAMi-J27 interaction follows quite closely the HF IV density.  This is the reason why the HF and PVC models give almost the same isospin impurity although the two models evaluate quite differently  the values \cite{HS,Xavi2}. Thus, the large isospin impurity obtained by the HF results takes into account implicitly 
the coupling between the single-particle wave functions and the collective IVGMR due to the Coulomb interaction.

The IV density of modified HF results extracted from Fig.~\ref{fig2a} is also shown in Fig.~\ref{fig:PVC-den} as SAMi-J27mod.  It is noted that the peak height of 
SAMi-J27mod is almost the same as those of HF and PVC results.  This feature encourages to extract the isospin impurity from the maximum of the experimental  IV density, which will be discussed hereafter.  

The CSB and CIB interactions were introduced in the context of Skyrme interactions in Refs.~\cite{Jacheck,Xavi,Sagawa}.
 While there are several possible channels in both CSB and CIB interactions, 
 we consider $ s $-wave interactions as the main channel, 
  \begin{subequations}
    \begin{align}
      v_{\rm{Sky}}^{\rm{CSB}} \left( \vec{r} \right)
      & =
        s_0
        \left( 1 + y_0 P_{\sigma} \right)
        \delta \left( \vec{r} \right)
        \frac{\tau_{z1} + \tau_{z2}}{4},
        \label{eq:Skyrme_CSB} \\
      v_{\rm{Sky}}^{\rm{CIB}} \left( \vec{r} \right)
      & =
        u_0
        \left( 1 + z_0 P_{\sigma} \right)
        \delta \left( \vec{r} \right)
          \frac{\tau_{z1} \tau_{z2}}{2}. 
        \label{eq:Skyrme_CIB}
    \end{align}
  \end{subequations}
 \vspace{-5mm}
 The energy density  of ISB part of the Skyrme interaction reads
 \vspace{-2mm}
  \begin{subequations}
    \begin{align}
      \mathcal{E}_{\rm{CSB}}
      & = 
        \frac{s_0 \left( 1 - y_0 \right)}{8}
        \left( \rho_n^2 - \rho_p^2 \right),  \\
      \mathcal{E}_{\rm{CIB}}
      & =
        \frac{u_0}{8}
        \left[
        \left( 1 - z_0 \right)
        \left( \rho_n^2 + \rho_p^2 \right)
        -
        2 \left( 2 + z_0 \right)
        \rho_n \rho_p
        \right].
    \end{align}
  \end{subequations}
The mean-field potentials of CSB and CIB can be evaluated by a functional derivative of energy density with respect to proton and neutron densities;
\begin{subequations} \label{CSB-EDF}
\begin{align}
V_{\rm CSB}^n(r)&=\frac{\delta \mathcal{E}_{\rm{CSB}}}{\delta \rho_n}=\frac{s_0(1-y_0)}{4}\rho_n,  \\
V_{\rm CSB}^p(r)&=\frac{\delta \mathcal{E}_{\rm{CSB}}}{\delta \rho_p}=-\frac{s_0(1-y_0)}{4}\rho_p,  
 \end{align}
  \end{subequations}
 and 
\begin{subequations} \label{CIB-EDF}
\begin{align}
V_{\rm CIB}^n(r)&=\frac{\delta \mathcal{E}_{\rm{CIB}}}{\delta \rho_n}=\frac{u_0(1-z_0)}{4}\rho_n -\frac{u_0}{4}(2+z_0)\rho_p, \\
V_{\rm CIB}^p(r)&=\frac{\delta \mathcal{E}_{\rm{CIB}}}{\delta \rho_p}=\frac{u_0(1-z_0)}{4}\rho_p  -\frac{u_0}{4}(2+z_0)\rho_n.   
 \end{align}
  \end{subequations}
In SAMi-ISB parameter sets \cite{Xavi}, the CSB and CIB interactions are optimized for a set of experimental data to be 
$s_0=-26.3 \, \mathrm{MeV} \, \mathrm{fm}^3$ and $u_0=25.8 \, \mathrm{MeV} \, \mathrm{fm}^3$ for the choice of spin-exchange parts $y_0=-1$ and $z_0=-1$.

\begin{figure}[htb]
\begin{center}
\includegraphics[clip,width=0.9\linewidth,bb=10 0 470 307]{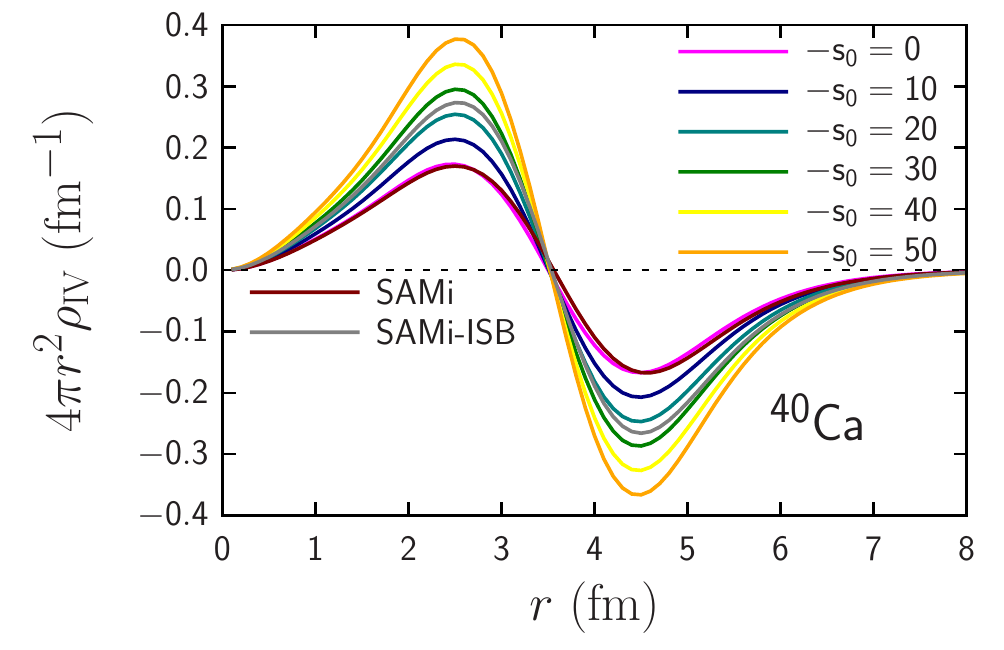}%
\caption{(color online) IV density in ${}^{40} \mathrm{Ca}$ calculated by a HF model with various SAMi parameter sets. 
The parameter $-s_0$ of CSB interaction is varied from $0$ to $50 \, \mathrm{MeV} \, \mathrm{fm}^3$
with a step of $ 10 \, \mathrm{MeV} \, \mathrm{fm}^3$ keeping CIB parameter $u_0=0 \, \mathrm{MeV} \, \mathrm{fm}^3$ on top of the central part of SAMi-ISB parameter set.  
See the text for details.
\label{fig:CIB-CSB}}
\end{center}
\vspace{-.5cm}
\end{figure}
 
We study how much the IV density is changed  by  different values of ISB interactions in Fig.~\ref{fig:CIB-CSB}.
 In Fig. \ref{fig:CIB-CSB}, the strength of 
 CSB interaction is varied from $ -s_0 = 0 $ to $ 50 \, \mathrm{MeV} \, \mathrm{fm}^3$ with a step of $ 10 \, \mathrm{MeV} \, \mathrm{fm}^3$,
 keeping CIB parameter $u_0=0 \, \mathrm{MeV} \, \mathrm{fm}^3$, as did in Ref.~\cite{Naito}.
 It is remarkable that the CSB effect enhances largely the IV density.  The CIB parameter dependence is also checked to change $u_0=0 $, $ 25 $, and $ 50 \,\mathrm{MeV} \, \mathrm{fm}^3$
keeping the CSB parameter  $-s_0=0 \,\mathrm{MeV} \, \mathrm{fm}^3$ as shown in Fig. 2 of the supplemental materials.  We found that the CIB interaction does not change at all the magnitude of IV density in contrast to the results of the change of CSB interaction strength (see for details the supplemental materials).
 
\begin{figure}[htb]
\begin{center}
\includegraphics[
width=0.9\linewidth,bb=0 0 482 312]{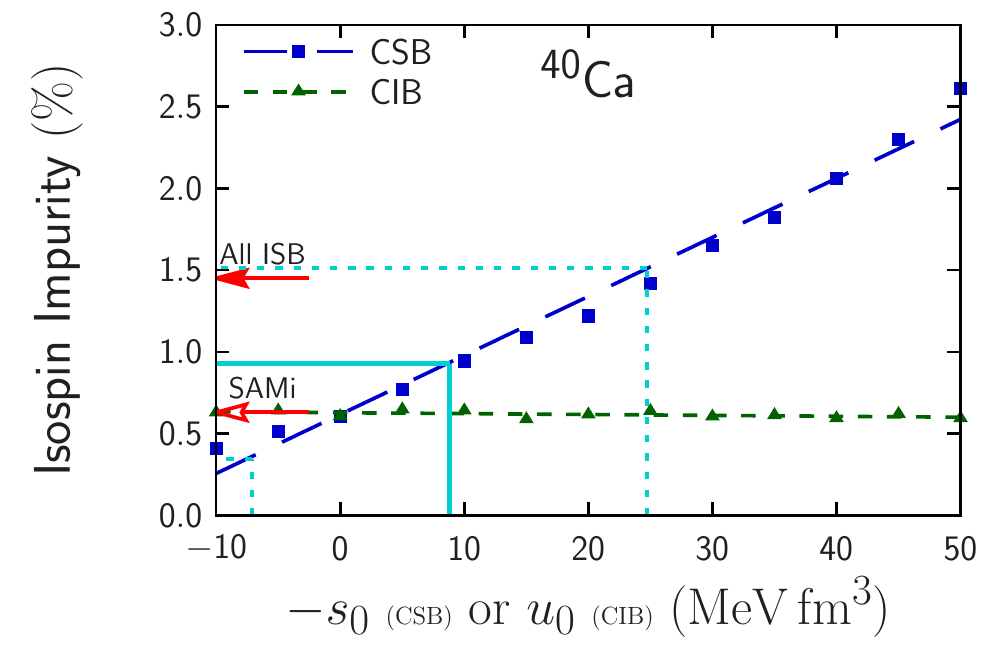}%
\vspace{-0.3cm}
\caption{(color online) Isospin impurity  in ${}^{40} \mathrm{Ca}$ calculated by a HF model  with various strength of CSB and CIB interactions.
  The strength of CSB, $-s_0$,   and CIB, $u_0$,  are varied from $-10$ to $50 \, \mathrm{MeV} \, \mathrm{fm}^3$ with a step of $ 5 \, \mathrm{MeV} \, \mathrm{fm}^3$.  The correlation coefficient between two values is $r=0.991$. The parameter set SAMi has no ISB interactions, while SAMi-ISB (denoted All ISB) has the CSB and CIB strength
 $-s_0=26.3 \, \mathrm{MeV} \, \mathrm{fm}^3$  and  $u_0=25.8 \, \mathrm{MeV} \, \mathrm{fm}^3$, respectively.  The cyan line is drawn to extract the CSB strength $-s_0$ from the isospin impurity determined in Fig.~\ref{fig:peak-impurity}.  The cyan dashed lines are experimental uncertainty. 
See the text for details.
\label{fig:impurity}}
\end{center}
\vspace{-0.7cm}
\end{figure}

Figure \ref{fig:impurity} shows the isospin impurity as a function of $-s_0 $ ($u_0$) for the CSB (CIB) interaction.  The value indicated by the arrow with ``SAMi'' is  induced entirely by the Coulomb interaction,  while the arrow labelled by  ``All ISB'' is the one by SAMi-ISB.
There is a clear difference between the CSB and CIB dependence.  Namely, the CIB has no effect on the isospin impurity, while the CSB gives a large enhancement on the isospin impurity. The correlation coefficient between the CSB strength and the isospin impurity is very high as $r=0.991$.  
 Consequently, the parameter set SAMi-ISB gives much larger value than that of SAMi, i.e., more than a factor 2  larger than the SAMi value.

This peculiar feature of CSB and CIB can be understood by studying the mean-field potentials originated by CSB and CIB interactions.
Taking $y_0=z_0=-1$ as is the same as SAMi-ISB \cite{Xavi},  the mean-field potential of CSB and CIB interactions are expressed as
 Eq.   \eqref{CSB-EDF} shows that  
the CSB contribution has a pure IV character to enhance the difference between neutron and proton density distributions. 
On the other hand, the CIB potentials \eqref{CIB-EDF} do not give any enhancement on the IV density.  
The CIB interaction violates in general the isospin invariance characterized by the   rotation in the isospin space. However, the CIB interaction holds the charge symmetry due to 
the isospin rotation by 180$^\circ$ about the 
$y$-axis \cite{Miller}.
Thus, the characteristic features of CSB and CIB interactions in Figs.~\ref{fig:CIB-CSB} and  \ref{fig:impurity}   are interpreted as the outcome of  the intrinsic nature of CSB and CIB interactions.

 \begin{figure}[t]
\begin{center}
\includegraphics[clip,width=0.95\linewidth,bb=0 0 482 312]{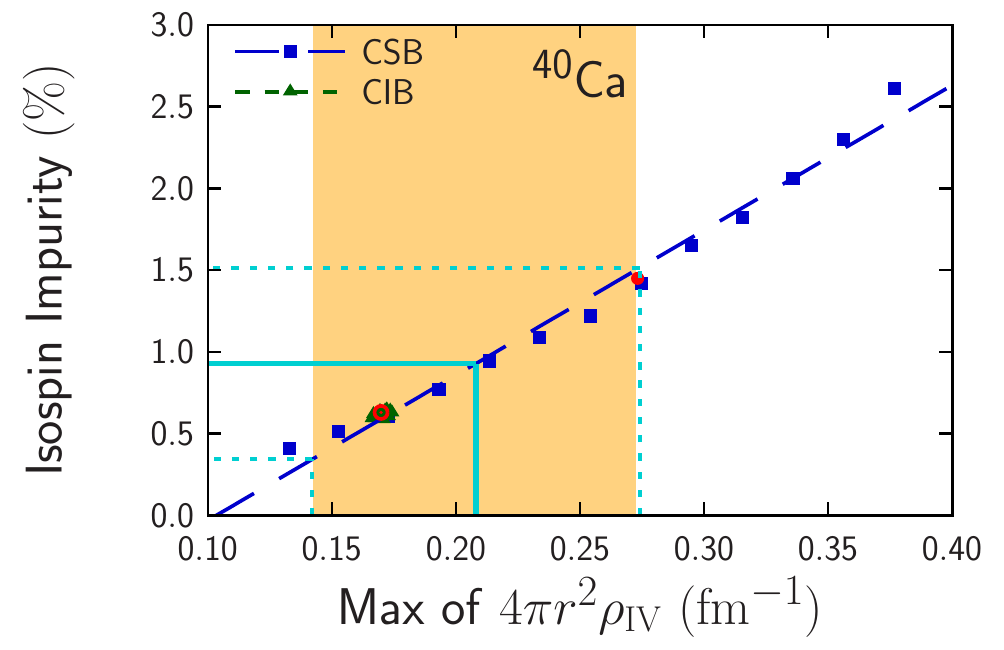}%
\caption{(color online) Isospin impurity vs.~the peak of IV density  in ${}^{40} \mathrm{Ca}$.  The calculations are performed by  changing the strength of 
CSB and CIB interactions on top of the central part of SAMi-ISB parameter set. The correlation coefficient between two values is $r=0.992$.  The red circle and the double red circle show the results of SAMi-ISB and SAMi  interactions, respectively.   The results of CIB interaction are seen  as one green  triangle since the results are  changed scarcely by the change of strength of CIB interaction and overlapped largely in the scale of this figure.}
\label{fig:peak-impurity}
\end{center}
\vspace{-0.5cm}
\end{figure}

Figure \ref{fig:peak-impurity} shows the correlation between the maximum of  IV density and the isospin impurity.  The correlation coefficient is very high as $r=0.992$.  This clear correlation is expected from very smooth increase of  both the IV density and the isospin impurity in Figs.~\ref{fig:CIB-CSB} and  \ref{fig:impurity}.  It could be possible to extract the isospin impurity from the peak height of IV density when both the proton and neutron densities are available experimentally.  The experimental peak height of IV density is shown to be $0.208\pm 0.066$ fm$^{-1}$ in Fig.~\ref{fig:PVC-den}.   The isospin impurity is then extracted from the correlation plot of 
Fig.~\ref{fig:peak-impurity}  as
\be
\varepsilon^2=0.928\pm 0.586.
\ee
This central value is about 50\% larger than the value of RPA  calculations without the ISB forces in Ref. \cite{HS}.  From this value of the isospin impurity, the strength of CSB interaction $s_0$ is further obtained  as 
\be
s_0=- \left(8.80 \pm 16.0 \right) \, \mathrm{MeV} \, \mathrm{fm}^3, 
\ee    
from the correlation plot in Fig.~\ref{fig:impurity}.

\begin{figure}[htb]
\begin{center}
\vspace{-0.8cm}
\includegraphics[
width=0.85\linewidth,bb=30 0 462 612]{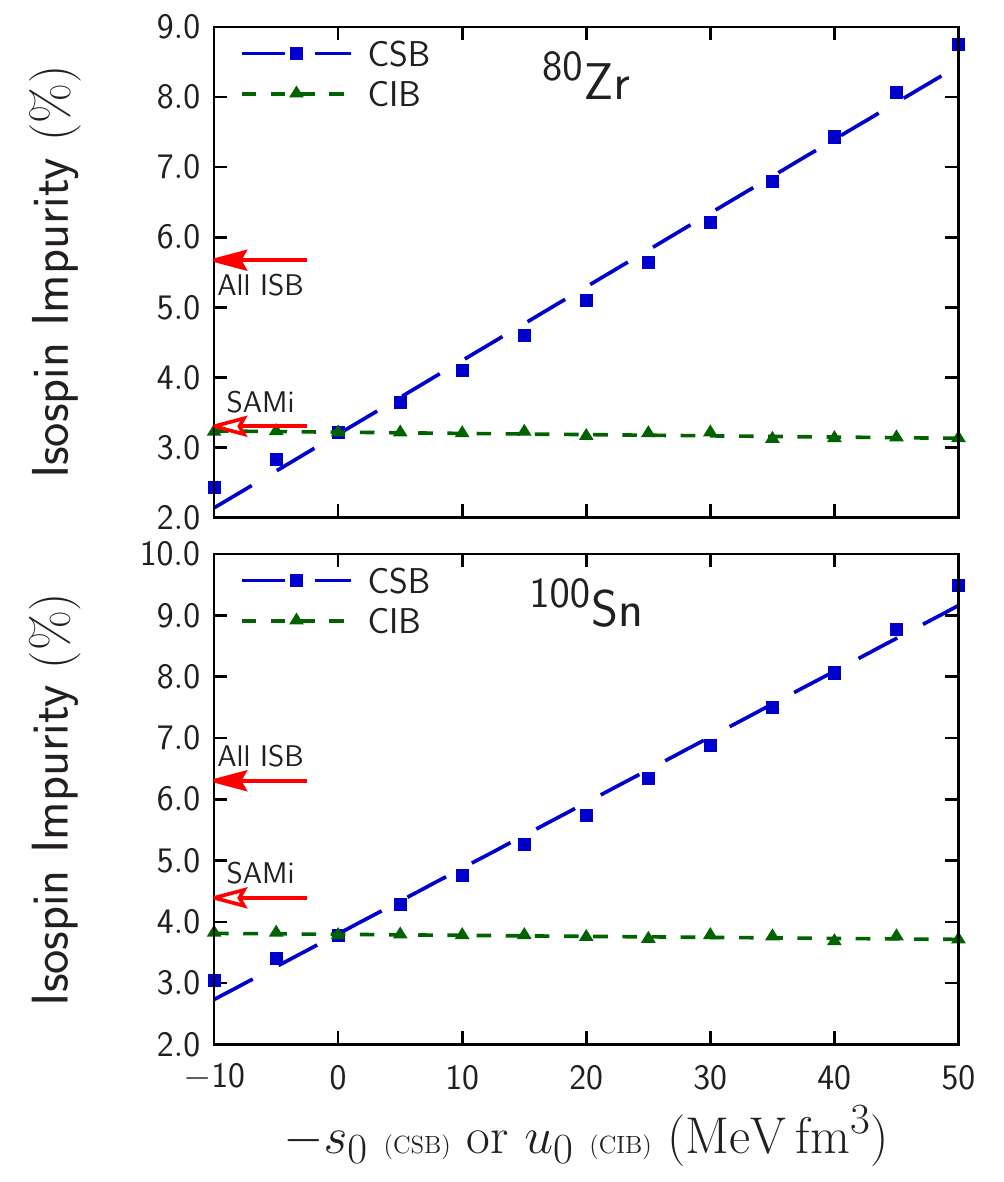}%
\vspace{-0.5cm}
\caption{(color online) Isospin impurities  in ${}^{80} \mathrm{Zr} $ and $ {}^{100} \mathrm{Sn} $ calculated by a HF model  with various strength of CSB and CIB interactions on top of the central part of SAMi-ISB interaction.   
The strength of CSB, $-s_0$,   and CIB, $u_0$,  are varied from $-10$ to $50 \, \mathrm{MeV} \, \mathrm{fm}^3$ with a step of $ 5 \, \mathrm{MeV} \, \mathrm{fm}^3$.  The correlation coefficient between the isospin impurity and the CSB interaction  is $r=0.996$. The parameter set SAMi has no ISB interactions, while SAMi-ISB (denoted All ISB) has the CSB and CIB strength
 $-s_0=26.3 \, \mathrm{MeV} \, \mathrm{fm}^3$  and  $u_0=25.8 \, \mathrm{MeV} \, \mathrm{fm}^3$, respectively.  
\label{fig:impurity-Zr-Sn}}
\end{center}
\vspace{-0.5cm}
\end{figure}

The calculated results of CSB dependence of isospin impurities in other  $N=Z$ nuclei, $^{80}$Zr and $^{100}$Sn, are shown in Fig. \ref{fig:impurity-Zr-Sn}.
We found again very similar strong CSB dependence of the impurities of $^{80}$Zr and $^{100}$Sn 
 in these figures, while the CIB interaction does not give any appreciable effect.

 It is  shown   that the correlation between the area of IV density and the isospin impurity is
as strong as that between the peak height and the isospin impurity shown in Fig.~\ref{fig:peak-impurity} (see  the supplemental materials for details).  These features support our procedure to extract the isospin impurity, as well as the CSB strength,   from  the correlation between the magnitude of IV density and the isospin impurity. 

\section{Summary and future perspectives}
In summary, we studied the  IV and IS densities of ${}^{40} \mathrm{Ca}$ by using the mean-field and PVC models.
 As the mean-field   models, we took 
 Skyrme SAMi-J model.  We found an  appreciable difference between the experimental and calculated IS densities in the interior part and also 
  dilute density region of ${}^{40} \mathrm{Ca}$. 
   This  difference  suggests  the modification of  density distribution by reduced occupation probabilities of single-particle states near the Fermi surface, which may be caused by  many-body correlations beyond the mean-field model.

    The IV density is  examined in the  Skyrme HF model, and also the PVC  model taking into account  the IV  GMR.   We found a close resemblance between the HF and PVC IV densities, which will cause the same amount of isospin impurity in the ground state of ${}^{40} \mathrm{Ca}$.   It was also shown  that the magnitude of IV density is changed largely by the CSB interaction, while the CIB interaction gives no appreciable effect.   It is  found the CSB interaction shows  a strong linear correlation with the maximum of IV density as well as 
  with  the isospin impurity.  Thus, the magnitude of IV density gives a good clue to determine experimentally the isospin impurity and the magnitude of CSB interaction.  This characteristic feature of IV density appear not only in ${}^{40} \mathrm{Ca}$, but also in other $N=Z$ nuclei, ${}^{80} \mathrm{Zr} $ and ${}^{100} \mathrm{Sn}$. 
   Precise measurements of the IV density is
   desperately desired to obtain experimental information of the isospin impurity and also the CSB interaction.  

This work was supported by  JSPS KAKENHI  Grant Numbers 15H054, JP19K03858, and JP19J20543.
 We thank M. Honma, N. Shimizu, and T. Fukui for informing us shell model results of $^{40}$Ca.  
The numerical calculations were partly performed on cluster computers at the RIKEN iTHEMS program.





\end{document}


\title{Isovector density and isospin impurity in $^{40} \mathrm{Ca} $\\
Supplemental materials}

 \author{H.~Sagawa\fnref{1,2}}
\address[1]{RIKEN Nishina Center, Wako, Saitama 351-0198, Japan}
\address[2]{Center for Mathematical Sciences, the University of Aizu, Aizu-Wakamatsu, Fukushima 965-8580, Japan}

\author{S.~Yoshida\fnref{3}}
\address[3]{Science Research Center, Hosei University,
2-17-1 Fujimi, Chiyoda, Tokyo 102-8160, Japan}

\author{T.~Naito\fnref{1,4}}
  \address[4]{Department of Physics, Graduate School of Science, The University of Tokyo, Tokyo 113-0033, Japan}

\author{T.~Uesaka\fnref{1,5}}
\address[5]{RIKEN Cluster for Pioneering Research, Wako, Saitama 351-0198, Japan}

 \author{J.~Zenihiro\fnref{1,6}}
\address[6]{Department of Physics, Kyoto University, Kitashirakawa-Oiwake, Sakyo, Kyoto 606-8502, Japan}

\author{J.~Tanaka\fnref{1}}
 \author{T. Suzuki\fnref{7}}
\address[7]{Department of Physics, College of Humanities and Sciences, Nihon University, Sakurajosui 3, Setagaya-ku, Tokyo 156-8550, Japan} 
\vspace{-1cm}

\maketitle
\section{Nuclear matter and ground-state properties calculated by using  RMF Lagrangians  and Skyrme parameters}
\begin{table*} [tb]
  \caption{Nuclear matter and ground-state properties calculated by using  RMF Lagrangians DDME-J family and Skyrme parameters. 
    $K_{\infty}, J, L$, and $K_{\mathrm{sym}}$ are 
the nuclear matter incompressibility, the volume, the slope and the second derivative terms of the symmetry energy, while $K_{\tau}$ and $m_{\mathrm{eff}}$ are the second derivative term of the incompressibility with respect to the asymmetry coefficient 
$\delta\equiv\rho_{\rm IV}/\rho_{\rm IS}$, 
and the effective mass (Dirac mass for RMF model) , respectively \cite{Yoshida2020}.
Calculated proton, neutron r.m.s. radii and the neutron-skin $\Delta r_{np}=r_n-r_p$ of ${}^{40} \mathrm{Ca}$ 
  are also listed. 
 \label{tab2}}
  \begin{tabular}{l|D{.}{.}{1}D{.}{.}{2}D{.}{.}{2}D{.}{.}{2}D{.}{.}{2}D{.}{.}{2}|D{.}{.}{3}D{.}{.}{3}D{.}{.}{3}}
    \hline
    Parameters & \multicolumn{1}{c}{$K_{\infty}$ (MeV)} & \multicolumn{1}{c}{$J$ (MeV)} & \multicolumn{1}{c}{$L$ (MeV)} &\multicolumn{1}{c}{$K_{\mathrm{sym}}$ (MeV)} & \multicolumn{1}{c}{$K_{\tau}$ (MeV)} & \multicolumn{1}{c|}{$m_{\mathrm{eff}}/m$} & \multicolumn{1}{c}{$r_p$ (fm)} & \multicolumn{1}{c}{$r_n$ (fm)} & \multicolumn{1}{c}{$\Delta r_{np}$ (fm)} 
    \\ \hline
    DDME-J30& 249.9&  30.06&   30.05&   -13.24&  -235.93&  0.617 & 3.365& 3.313 & -0.052 \\
    DDME-J32& 250.0&  31.97&   46.44&   -83.75&  -424.27&  0.619 & 3.369& 3.317 & -0.052 \\
    DDME-J34& 249.9&  33.97&   61.98&  -106.22&  -556.87&  0.619 & 3.368& 3.317 & -0.051 \\
    DDME-J36& 249.8&  35.99&   85.39&   -76.28&  -697.55&  0.619 & 3.367& 3.317 & -0.050 \\
    DDME-J38& 250.0&  38.03&  110.74&    10.86&  -799.28&  0.620 & 3.372& 3.323 & -0.049 \\ \hline
    SAMi-J27& 245.0&  27.00&   30.00&  -158.04&  -296.55&  0.675 & 3.391& 3.344& -0.047 \\
    SAMi-J29& 245.0&  29.00&   51.60&  -100.17&  -338.86&  0.675 & 3.389& 3.342& -0.047 \\
    SAMi-J31& 245.0&  31.00&   74.37&   -37.35&  -382.14&  0.675 & 3.387& 3.341& -0.046 \\
    SAMi-J33& 245.0&  33.00&   95.41&    19.64&  -423.93&  0.675 & 3.384& 3.338& -0.046 \\
    SAMi-J35& 245.0&  35.00&  114.95&    71.64&  -464.42&  0.675 & 3.378& 3.334& -0.044 \\ 
  \end{tabular}
\end{table*}
Table 1 shows nuclear matter and rms radii calculated by using  RMF Lagrangians  DDME-J family  and Skyrme parameters  SAMi-J 
family.

\begin{figure}[tb]
\begin{center}
 \includegraphics[clip,width=1.0\linewidth,bb=0 250 550 750]{
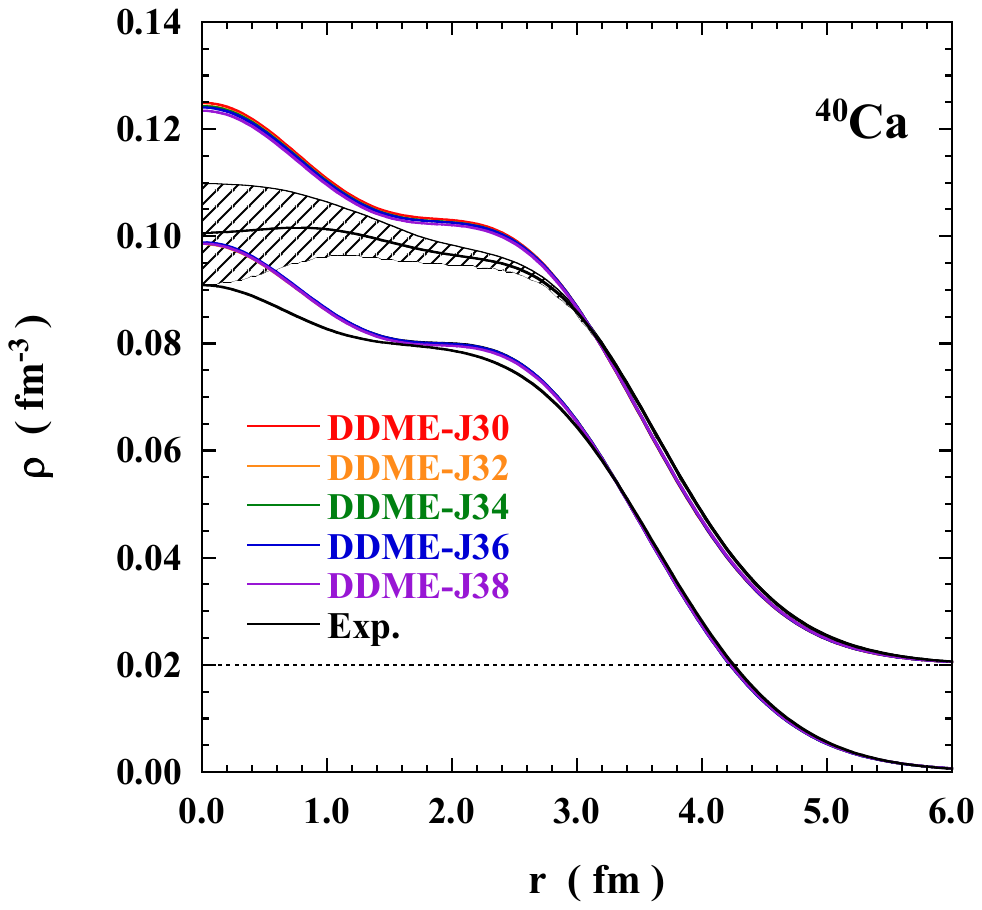}
\caption{(color online) Experimental proton and neutron densities of ${}^{40} \mathrm{Ca}$ together with calculated ones using RMF  DDME-J interactions. 
For a guide to eyes, the neutron density is shifted by 0.02 fm$^{-3}$.  
  The black solid lines show experimental data  taken from Ref.~\cite{electron-sca} for protons and from Ref.~\cite{Zenihiro} for neutrons. 
 The shaded area of experimental neutron density shows experimental uncertainties of statistical and systematic errors.
}
 \label{fig1}
\end{center}
\end{figure}

Figure  \ref{fig1} shows the calculated neutron and proton densities of $^{40}$Ca by using  RMF Lagrangians 
DDME-J family.   The experimental proton and neutron densities are also shown in Figure  \ref{fig1}.  
  The DDME-J family has  a variation in the symmetry energy coefficient $J$ from $30$ to $38 \, \mathrm{MeV}$ as well as other  symmetry energy coefficients $L$ and $K_{\mathrm{sym}}$ as listed in Table I. 
We can see that  the Lagrangian with smaller J-value gives a largely  enhanced shoulder of neutron density  at $r = 2.6 \, \mathrm{fm}. $ 
  The shoulder height is produced by DDME-J32 at the best.  Lagrangian dependence of proton shoulder at $r=$2.3 fm is not clearly seen in Fig.~\ref{fig1}. 

\begin{figure}[htb]
\begin{center}
\includegraphics[clip,width=1.0\linewidth,bb=0 0 482 312]{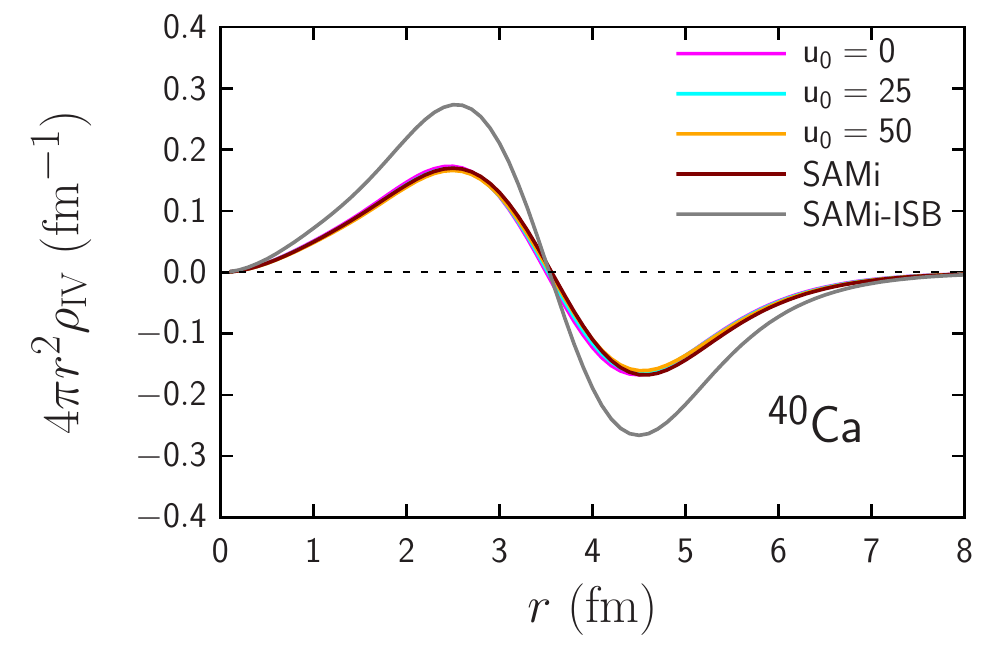}%
\caption{(color online)  IV density in ${}^{40} \mathrm{Ca}$ calculated by a HF model with various SAMi parameter sets. 
The parameter $u_0$ of CIB interaction is varied from $0$ to $50 \, \mathrm{MeV} \, \mathrm{fm}^3$
with a step of $ 25 \, \mathrm{MeV} \, \mathrm{fm}^3$ keeping CSB parameter $s_0=0 \, \mathrm{MeV} \, \mathrm{fm}^3$ on top of the central part of SAMi-ISB parameter set.  
See the text for details.
\label{fig:IV-CIB}}
\end{center}
\end{figure}

\begin{figure}[htb]
\begin{center}
\includegraphics[clip,width=1.0\linewidth,bb=0 0 482 312]{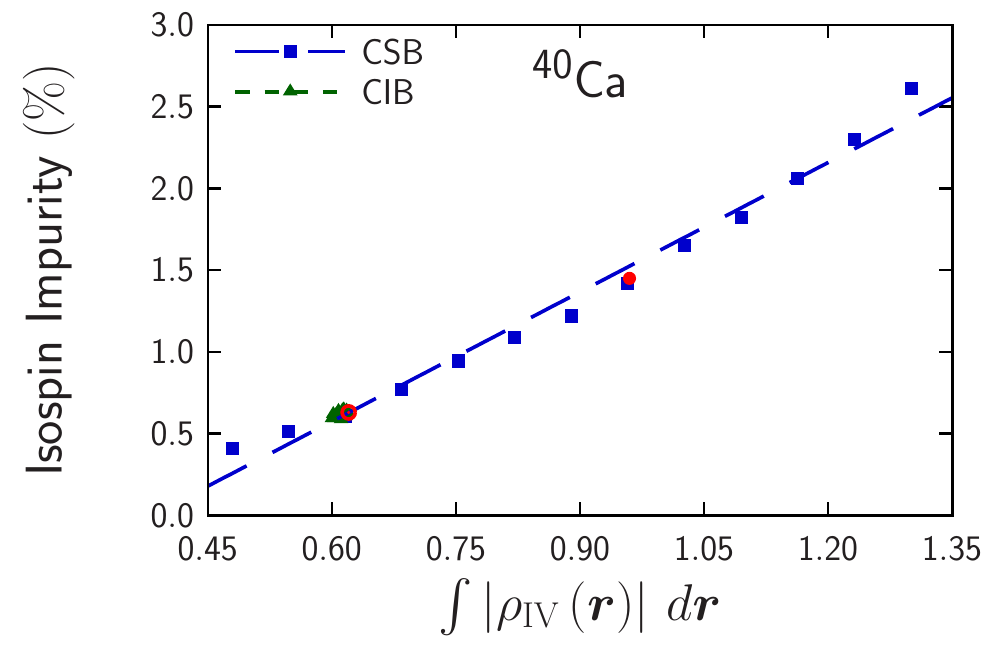}%
\caption{(color online) Isospin impurity vs.~the area of IV density  in ${}^{40} \mathrm{Ca}$.  The calculations are performed by  changing the strength of 
CSB and CIB interactions on top of the central part of SAMi-ISB parameter set. The correlation coefficient between two values is $r=0.991$.  The red circle and the double red circle show the results of SAMi-ISB and SAMi  interactions, respectively.   
The results of CIB interaction are seen  as one green  triangle since the results are  changed scarcely by the change of strength of CIB interaction and overlapped largely in the scale of this figure.} 
\label{fig:area-impurity}
\end{center}
\end{figure}

\section{The correlation between the area of IV density and the isospin impurity in ${}^{40} \mathrm{Ca}$}

Figure \ref{fig:IV-CIB} shows the CIB interaction dependence of IV density in ${}^{40} \mathrm{Ca}$ calculated by a HF model with various SAMi parameter sets.  As is clearly seen in this figure, the IV density gets essentially no effect by the CIB interaction because of the reason mentioned in the main text.

Figure \ref{fig:area-impurity} shows the correlation between the area of IV density and the isospin impurity in ${}^{40} \mathrm{Ca}$.  
The calculations are performed by  changing the strength of 
CSB and CIB interactions on top of the central part of SAMi-ISB parameter set.  The correlation between the area and the isospin impurity 
is as strong as that between the peak height of IV density and the isospin impurity shown in Fig.~6 in the main text.

\section{Particle occupation numbers of $sd$-$pf$ shell orbits in $^{40}$Ca}
Skyrme HF-Bogolyubov calculations  are performed with SAMi interaction and the volume-type and mixed-type pairing interactions with the code HFBTHO. 
We found that there are essentially no occupation probabilities $v_j^2$ in $pf$-shell orbits.  The results are listed in Table II.
Large scale shell model calculations have been performed including many-particle many-hole excitations from $sd$-shell to $pf$-shell configurations.
  The particle occupation numbers are summarized in Table \ref{tabPOP}.  In Ref. \cite{Brown}, the active model space is $(1d_{3/2}, 1f_{7/2}, 2p_{3/2})$, while 
  the $1d_{3/2}, 2s_{1/2}$ orbits are also included in Ref. \cite{Shimizu}.  We perform also shell model calculations with the configurations of  2-particle 2-hole (2p-2h) and 4p-4h excitations from the closed shell core of $^{40}$Ca.   The full $sd$ and $pf$ shell orbits  are involved with $sdpf$-mu shell model effective interactions.  The experimental occupation numbers are also listed obtained from proton transfer reactions \cite{p-trans}, and also from the analysis of   neutron density distributions \cite{main-text}.

\begin{table*} [t]
  \caption{Particle occupation numbers, $(2j+1)v_j^2$, in $sd$-shell to $pf$ shell configurations in $^{40}$Ca.  The column with a bar (---) is not involved in the shell model calculations.
  Experimental data of proton transfer reaction are taken from Ref. \cite{p-trans}.  
\label{tabPOP}}
\begin{center}
  \begin{tabular}{l|ccc|cccc}
    \hline
   Model & $1d_{5/2}$ & $2s_{1/2}$ & $1d_{3/2}$ &  $1f_{7/2}$ &  $2p_{3/2}$ &$2p_{1/2}$ & $1f_{5/2}$   \\\hline 
HFB (SAMi)  & 6.00  &2.00  &4.00  &   0.00 & 0.00 & 0.00 & 0.00  \\  
 $dpf$-shell \cite{Brown} & --- & --- & 3.30 & 0.63 & 0.07 & ---  & --- \\
  $sdpf$-msd4 \cite{Shimizu} & 5.902 & 1.908 & 3.477 & 0.617 & 0.096 & --- & --- \\
  $sdpf$-mu $(2p$-$2h)$ &  5.864 & 1.933 & 3.845 & 0.191 & 0.040 & 0.020 & 0.107 \\
  $sdpf$-mu $(4p$-$4h)$    &  5.727& 1.854 & 3.660 & 0.421 & 0.089 & 0.041 & 0.208 \\
  exp. (p-transfer) \cite{p-trans} & 6.0 & 1.70& 3.59 & 0.56  & 0.15 &  &  \\
  exp. (neutron density) \cite{main-text}& 6.0   & 1.342  & 3.431 &  1.227 & &  &  
  \end{tabular}
 \end{center}
 \end{table*}

\section{Isovector densities and the isospin impurities of ${}^{80} \mathrm{Zr} $ and $ {}^{100} \mathrm{Sn} $}

\begin{figure}[t]
\begin{center}
\vspace{-5.0cm}
\includegraphics[clip,width=1.0\linewidth,bb=0 0 482 567]{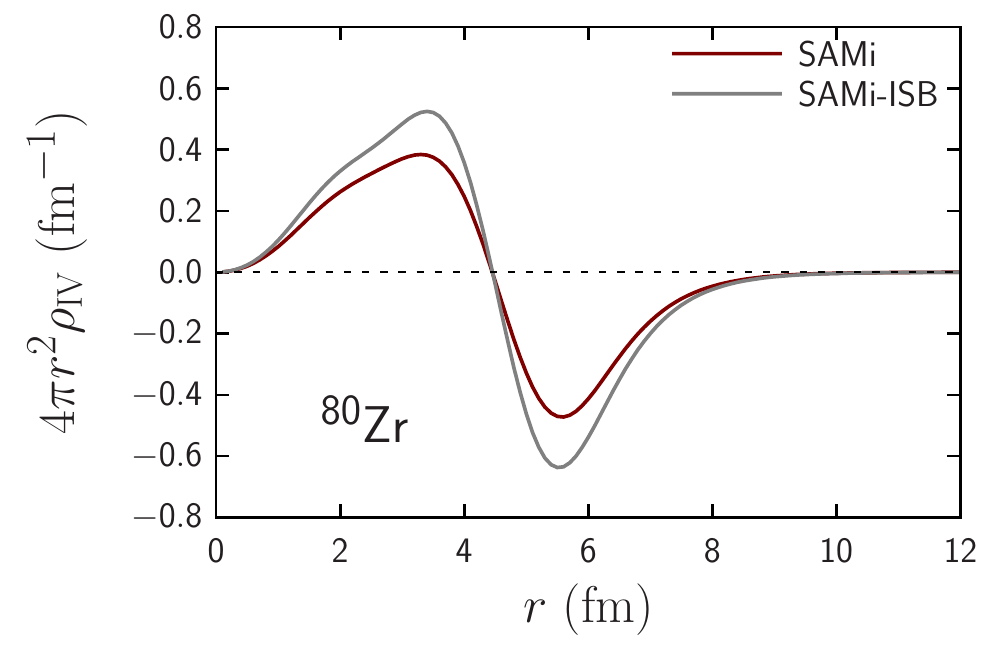}%
\caption{(color online) IV density in ${}^{80} \mathrm{Zr} $ calculated by a HF model with  SAMi and SAMi-ISB interactions, respectively.  \label{fig:IVden-Zr}}
\end{center}
\end{figure}

\begin{figure}[htp]
\begin{center}
\vspace{-5.0cm}
\includegraphics[clip,width=1.0\linewidth,bb=0 0 482 567]{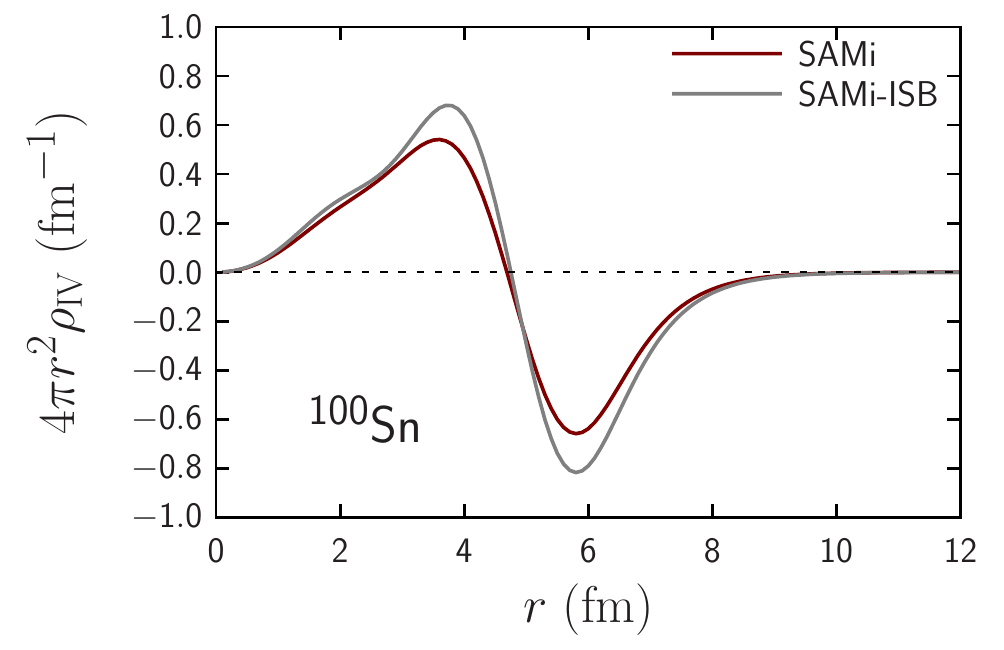}%
\caption{(color online) IV density in $ {}^{100} \mathrm{Sn} $ calculated by a HF model with  SAMi and SAMi-ISB interactions, respectively.  \label{fig:IVden-Sn}}
\end{center}
\end{figure}

\begin{figure}[htb]
\begin{center}
\includegraphics[clip,width=1.0\linewidth,bb=0 0 482 312]{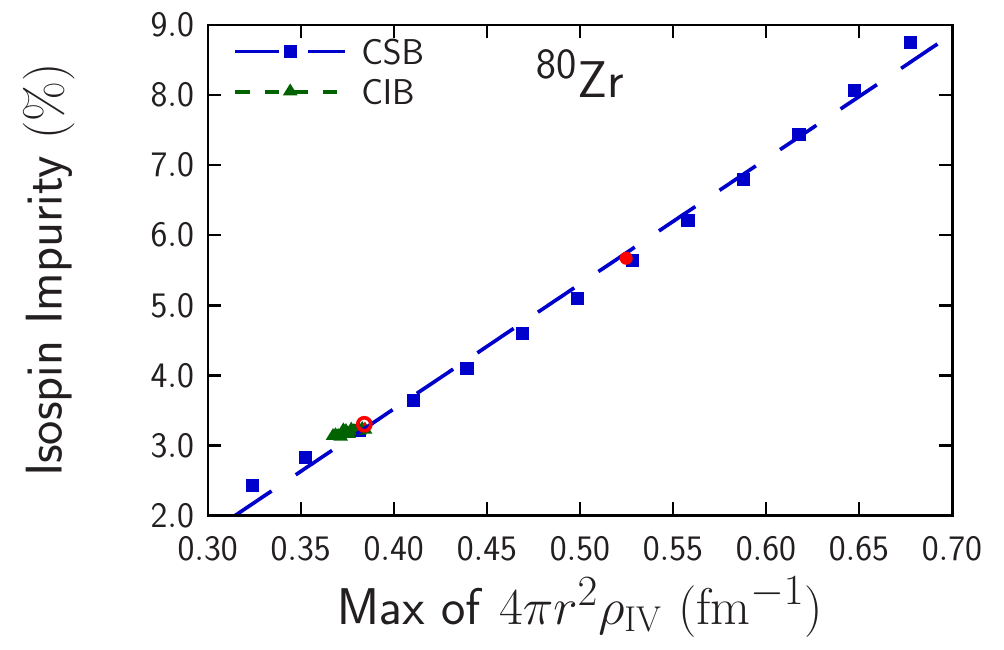}%
\caption{(color online) Isospin impurity vs.~the peak height of IV density  in ${}^{80} \mathrm{Zr} $.  The calculations are performed by  changing the strength of 
CSB and CIB interactions on top of the central part of SAMi-ISB parameter set. The correlation coefficient between two values is $r=0.997$.  The red circle and the double red circle show the results of SAMi-ISB and SAMi  interactions, respectively.   The results of CIB interaction are seen  as one green  triangle since the results are  changed scarcely by the change of strength of CIB interaction and overlapped largely in the scale of this figure.}
\label{fig:peak-impurity-Zr}
\end{center}
\end{figure}

\begin{figure}[htb]
\begin{center}
\includegraphics[clip,width=1.0\linewidth,bb=0 0 482 312]{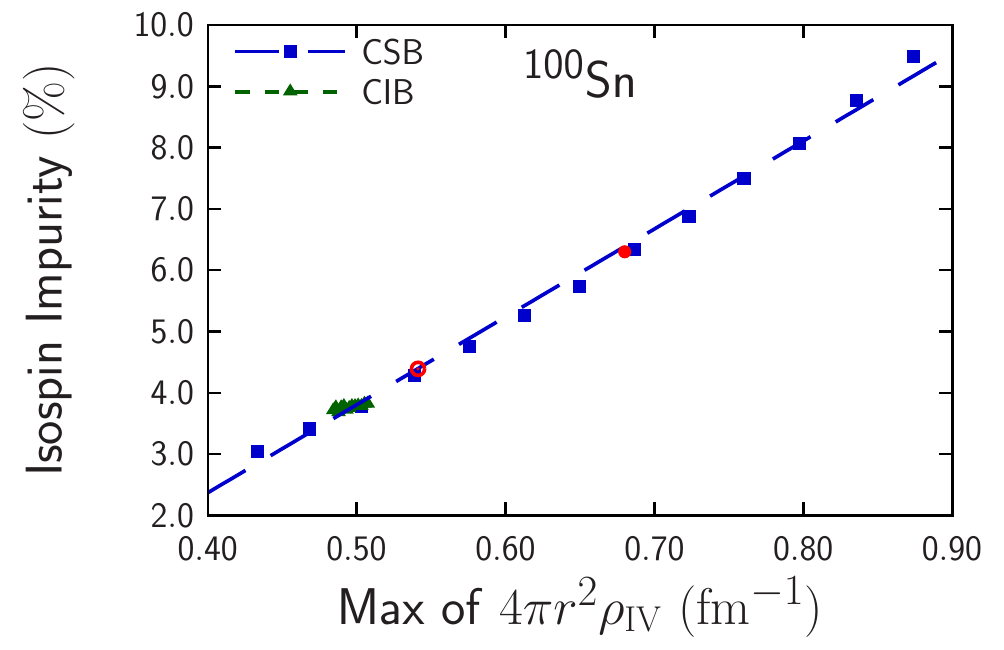}%
\caption{(color online) The same as Fig. \ref{fig:peak-impurity-Zr}, but for $ {}^{100} \mathrm{Sn} $. The correlation coefficient between two values is $r=0.998$.
\label{fig:peak-impurity-Sn}}
\end{center}
\end{figure}

 The isovector densities and the isospin impurities of $N=Z$ nuclei, ${}^{80} \mathrm{Zr} $ and $ {}^{100} \mathrm{Sn} $,  are shown  in terms of the charge symmetry breaking (CSB) and the charge independence breaking (CIB) interactions.  The HF IV densities calculated with SAMi and SAMi-ISB Skyrme interactions are shown 
in Figs. \ref{fig:IVden-Zr} and \ref{fig:IVden-Sn}, respectively.  It is well recognized that the IV density is enlarged by the ISB interactions in the SAMi-ISB parameter set.  The 
radial dependence is close to the Werntz-\"Uberall-type transition density with a small variation in the internal part due to the shell effect. 
 The isospin impurities  in ${}^{80} \mathrm{Zr} $ and $ {}^{100} \mathrm{Sn} $ calculated by a HF model  with various strength of CSB and CIB interactions on top of the central part of SAMi-ISB Skyrme interaction are shown in the main text. 
The correlation between the isospin impurity and the maximum of IV density is shown in Fig.~\ref{fig:peak-impurity-Zr} for ${}^{80} \mathrm{Zr} $  and
in Fig.~\ref{fig:peak-impurity-Sn}  for $ {}^{100} \mathrm{Sn} $.  
 The correlation coefficients between the isospin impurity and the maximum of IV density are  $r=0.997$ for ${}^{80} \mathrm{Zr} $ in Fig.~\ref{fig:peak-impurity-Zr}  and  $r=0.998$ for $ {}^{100} \mathrm{Sn} $ in Fig.~\ref{fig:peak-impurity-Sn}, respectively

